\documentclass[runningheads]{llncs}
\usepackage[T1]{fontenc}
\usepackage{hyperref}
\usepackage{graphicx}
\usepackage{xcolor}
\usepackage{todonotes}
\usepackage{orcidlink}
\usepackage{comment}
\usepackage{tikz}
\usepackage[absolute,overlay]{textpos}

\begin{document}

\title{SynthGuard: Redefining Synthetic Data Generation with a Scalable and Privacy-Preserving Workflow Framework
\thanks{\scriptsize 
Funded by the European Union (TEADAL 101070186 and LAGO 101073951). Views and opinions expressed are, however, those of the author(s) only and do not necessarily reflect those of the European Union. Neither the European Union nor the granting authority can be held responsible for them. Also supported by the Estonian Centre of Excellence in AI (EXAI), funded by the Estonian Ministry of Education and Research.}
}

\titlerunning{SynthGuard}

\author{Eduardo Brito\inst{1,2}\orcidlink{0009-0002-9996-6333} \and
Mahmoud Shoush\inst{1}\orcidlink{0000-0002-7423-9909} \and
Kristian Tamm\inst{1,2}\orcidlink{0009-0001-4500-9872} \and
Paula Etti\inst{1}\orcidlink{0000-0002-0701-9383} \and
Liina Kamm\inst{1}\orcidlink{0000-0003-1479-2195}}
\authorrunning{E. Brito et al.}

\institute{Cybernetica AS, Tallinn, Estonia \\
\email{\{eduardo.brito, mahmoud.shoush, kristian.tamm, paula.etti, liina.kamm\}@cyber.ee}
\and
University of Tartu, Tartu, Estonia
}

\maketitle              % typeset the header of the contribution

\begin{tikzpicture}[remember picture, overlay]
  \node[anchor=south, yshift=1cm] at (current page.south) {
    \fbox{%
      \begin{minipage}{0.9\textwidth}
        \footnotesize
        \textit{This is the extended version of the paper to appear in the Proceedings of the 1st International Workshop on Responsible Data Governance, Privacy, and Digital Transformation (RDGPT 2025), held in conjunction with the 20th International Conference on Availability, Reliability and Security (ARES 2025).}
      \end{minipage}
    }
  };
\end{tikzpicture}

\begin{abstract}

The growing reliance on data-driven applications in sectors such as healthcare, finance, and law enforcement underscores the need for secure, privacy-preserving, and scalable mechanisms for data generation and sharing. Synthetic data generation (SDG) has emerged as a promising approach but often relies on centralized or external processing, raising concerns about data sovereignty, domain ownership, and compliance with evolving regulatory standards. To overcome these issues, we introduce SynthGuard, a framework designed to ensure computational governance by enabling data owners to maintain control over SDG workflows. SynthGuard supports modular and privacy-preserving workflows, ensuring secure, auditable, and reproducible execution across diverse environments. In this paper, we demonstrate how SynthGuard addresses the complexities at the intersection of domain-specific needs and scalable SDG by aligning with requirements for data sovereignty and regulatory compliance. Developed iteratively with domain expert input, SynthGuard has been validated through real-world use cases, demonstrating its ability to balance security, privacy, and scalability while ensuring compliance. The evaluation confirms its effectiveness in implementing and executing SDG workflows and integrating privacy and utility assessments across various computational environments.

%Through real-world use cases, we show SynthGuard's ability to balance security, privacy, and scalability, meeting the demand for solutions to generate and share synthetic data.

\keywords{synthetic data generation \and privacy-preserving workflows \and data sovereignty and governance.}
\end{abstract}
\section{Introduction}\label{sec:introduction}

In today's data-driven landscape, sectors such as healthcare~\cite{Kononenko01,sidey2019machine}, finance~\cite{dixon2020machine}, and law enforcement~\cite{daskal2015law} face stringent legal and ethical challenges in managing and exchanging data. While artificial intelligence (AI), particularly machine learning (ML) and deep learning (DL), depends heavily on access to high-quality data to enhance operations and decision-making~\cite{jordan2015machine,sengupta2020review,GillXOPBSGSWASM22}, privacy concerns and regulatory frameworks such as the EU General Data Protection Regulation (GDPR) or the Directive (EU) 2016/860~\cite{regulation2020art,directive2016680} impose significant barriers to data availability. These restrictions result in lengthy compliance processes, stalling progress in data-driven innovation~\cite{abs-1710-08874,tolas2024gemsyd}.

Privacy-enhancing technologies (PETs) are helping overcome these barriers, allowing data processing, analysis, and insight extraction while safeguarding personal or commercially sensitive information~\cite{wagner2020privacy,KammBBO23,Majeed23}. Among these, synthetic data generation (SDG) has emerged as a promising solution that allows organizations to use representative synthetic data for research and development without risking individual privacy~\cite{ParkMGJPK18,taub2018differential,el2020practical,nikolenko2021synthetic}. However, existing SDG methods often involve external access to sensitive data, introducing privacy and security risks~\cite{Etti2023}. Furthermore, most research has focused on improving data generation methods or evaluating the utility and privacy of synthetic data~\cite{OsorioMarulandaEHIRI24}, neglecting the needs and challenges of implementing secure, scalable, and adaptable SDG workflows, while maintaining data sovereignty and adhering to evolving legal and ethical standards~\cite{Etti2023,hummel2021data}. 

Addressing these limitations may require new approaches that go beyond data generation mechanisms to tackle the operational and architectural challenges of large scale SDG workflows. Hence, we address the problem of how to %Such solutions must
enable data owners to maintain control over sensitive data and workflows while ensuring compliance with privacy regulations, supporting modularity and composability to meet diverse operational needs across heterogeneous environments. Transparency, auditability, and reliability are also essential to foster trust, ensure compliance, and enable practical deployment. 

To address this problem, %To meet these needs, 
we propose an approach that integrates governance, scalability, and compliance into SDG workflows. Particularly, in this paper, we:
\begin{enumerate}
    \item Present an architectural model for SDG workflows that prioritizes domain ownership, computational governance, and data sovereignty, ensuring that sensitive data and computations remain securely under the data owner’s control.
    \item Realize this architectural model into SynthGuard\footnote{\url{https://github.com/SynthGuard}}, a framework for modular and composable SDG workflow design that enables scalable and adaptable deployment across diverse operational environments.
    \item Validate SynthGuard through practical implementation in two European Union (EU) research projects, addressing legal and technical requirements of synthetic data sharing across six distinct use cases.
\end{enumerate}

SynthGuard was iteratively developed with domain expert input within the two EU research projects, covering diverse use cases that highlighted the challenges and needs of SDG across domains. Structured requirement elicitation ensured alignment with legal, technical, and operational constraints and validation confirmed its effectiveness in implementing and executing SDG workflows, while integrating privacy and utility assessments across different computational environments. Our results demonstrate that the framework maintains data sovereignty, while enabling flexible and efficient SDG execution, and preserving privacy and utility requirements.

The rest of this paper is organized as follows. First, in Section~\ref{sec:background}, we present the background. Then, in Section~\ref{sec:approach}, we start by outlining the requirements gathered from real-world use cases in two large-scale EU research projects, and follow with a discussion on the architectural principles needed to fulfil them, guiding SynthGuard development. In Section~\ref{sec:synthguard}, we detail the technical implementation of SynthGuard and, in Section~\ref{sec:validation}, we summarize the validation. Finally, we conclude the paper in Section~\ref{sec:conclusion}, suggesting future directions.

\section{Background}\label{sec:background}

The increasing adoption of data-driven services, particularly in the EU, underscores the need for secure data-sharing practices~\cite{KammBBO23}. However, ensuring privacy and regulatory compliance remains a significant challenge~\cite{regulation2020art,EU2022,EU2023}. SDG offers a promising approach to balancing privacy, data utility, and secure data sharing by creating data with realistic statistical properties while mitigating re-identification risks~\cite{taub2018differential,ParkMGJPK18,OsorioMarulandaEHIRI24}. Unlike traditional techniques such as anonymization, which modify real data points and often carry high re-identification risks~\cite{ChandraSTKVR22,abs-2405-20959}, SDG creates new data points that closely approximate real data, offering greater privacy guarantees~\cite{PingSH17}.

Research on SDG primarily focuses on developing methods to create synthetic datasets while assessing their utility and privacy~\cite{dankar2021fake,el2020practical}. Synthetic data can be categorized into \textit{realistic synthetic data} (RSD), \textit{causal synthetic data} (CSD), \textit{artificial synthetic data} (ASD), and \textit{hybrid synthetic data} (HSD)~\cite{HuWLLGGDFLS24,abs-2407-03672}. RSD replicates the statistical properties of original data using ML or DL, while CSD preserves causal relationships between data points~\cite{StoianDCLG24,ShiWTN22}. ASD relies on predefined rules or aggregated statistics, making it suitable when access to raw data is restricted~\cite{SoltanaSB17,LenattiPOFM23}. HSD combines these approaches, offering flexibility across domains~\cite{el2020practical}. Utility evaluations measure the effectiveness of synthetic data in substituting real data, often using domain-specific or statistical assessments~\cite{HittmeirEM19}. Generic metrics, such as univariate, bivariate, and population assessments, ensure synthetic data reflects the statistical characteristics of the original data~\cite{ChandraSTKVR22}. Privacy evaluations ensure compliance with legal frameworks by addressing risks of attribute disclosure, where individual characteristics might be inferred, or risks of identity disclosure, where individuals might be uniquely identified~\cite{OsorioMarulandaEHIRI24,KwatraT23,el2020practical}. Unlike data masking, which modifies real records while preserving their structure, these SDG methods generate new, independent data and are recognized as valid forms of synthetic data generation~\cite{el2020practical,HuWLLGGDFLS24}.

Despite advancements in new methods and evaluations, typical SDG processes often require data transfer or external access, introducing further privacy risks. Data availability and exchange are often constrained by legal, technical, and ethical concerns~\cite{Etti2023}. Sectors like healthcare, finance, and law enforcement need to manage vast amounts of sensitive data but face significant restrictions on reuse due to legal uncertainties~\cite{singh2021rise}. Key challenges range from obtaining necessary permissions for processing original data to addressing the legal status of synthetic data, whether pseudonymized or anonymized~\cite{Etti2023}. Furthermore, synthetic data processes must ensure compliance with evolving regulatory frameworks, balancing privacy and utility, which varies by use case~\cite{Case_T-557/20}. Two issues emerge from these challenges: limitations on data reuse, and lack of sufficient data for certain processing tasks. To our knowledge, no previous work has addressed all these needs, spanning legal, ethical, and technical dimensions, which necessitates rethinking how data is managed, shared, and processed.

To address these multifaceted constraints, emerging paradigms such as Data Mesh offer a promising foundation for rearchitecting synthetic data workflows. Originally proposed to manage decentralized data ownership and governance~\cite{machado2022data}, it advocates for a shift from centralized, monolithic data platforms to domain-oriented architectures and distributed data management. In data-intensive sectors, heterogeneous data pipelines are increasingly critical for handling diverse computational tasks and scaling to meet the growing demand for new data~\cite{tatineni2021ai}. The Data Mesh approach introduces key principles that may help balance scalability and security in (synthetic) data generation workflows~\cite{machado2022data,hummel2021data}: domain ownership, which establishes clear definitions and boundaries for what data is, where it resides, where it is processed, and by whom; computational governance, ensuring data is securely handled within environments governed by regulatory and operational constraints; and data sovereignty, emphasizing that sensitive data and computations should remain under the control of the data owner. In this paper, the terms (data) workflows and pipelines are used interchangeably. Conceptually, workflows represent the sequence of tasks involved in defining and building data processes, while pipelines refer to their specific implementation or execution~\cite{dessalk2020scalable,tatineni2021ai}.

\section{Requirements and Architectural Approach} \label{sec:approach}

This section presents the requirements and architectural approach that guided the design of SynthGuard. It begins by detailing the structured elicitation process used to identify key needs across use cases. Based on these requirements, we outline the architectural model developed to ensure privacy, scalability, and governance in SDG workflows. The following sections describe the technical implementation of this model in the SynthGuard framework and its validation against the identified requirements.

\subsection{Context and Requirements Gathering}

Our work was carried out within two Horizon Europe research projects: \textbf{LAGO}\footnote{\url{https://lago-europe.eu/}} and \textbf{TEADAL}\footnote{\url{https://teadal.eu/}}, each encompassing diverse SDG-related use cases:

\begin{itemize} 

\item \textbf{LAGO}: Aimed to enable \textbf{law enforcement agencies} to securely share data for research, training, and testing, in line with regulatory requirements. 

\item \textbf{TEADAL}: Included several domains: 

\begin{itemize} 

\item \textbf{Evidence-Based Medicine}: Secure patient data sharing for medical research. 

\item \textbf{Financial Governance}: Data aggregation across banking jurisdictions. 

\item \textbf{Smart Viticulture}: Business-critical data exchange via a digital platform. 

\item \textbf{Mobility}: Synthetic data generation for public use. 

\item \textbf{Regional Planning}: Cross-sector data sharing among public and private stakeholders. 

\end{itemize} 

\end{itemize}

To guide the framework design, we applied a structured requirements elicitation process. This included workshops and semi-structured interviews with domain experts, capturing operational contexts, regulatory constraints, and synthetic data needs. A use case analysis mapped data-sharing workflows, types of required data, and associated constraints, highlighting bottlenecks and opportunities for SDG to enhance compliance and security.

We also conducted surveys targeting researchers and practitioners, combining qualitative and quantitative questions on adoption barriers, preferred SDG techniques, and privacy/utility safeguards. Responses were analyzed to identify trends and gaps. In parallel, we performed a regulatory and standards review, analyzing legal frameworks such as the GDPR and the Data Governance Act, and sector-specific policies. Legal experts contributed to interpreting these regulations and embedding them into the framework. Full details of these activities are documented in the official public deliverables of LAGO and TEADAL.

\begin{sloppypar} 

This methodology yielded the requirements summarized in Table~\ref{table:requirements-elicited}. Common to all use cases are \textbf{ALL\_R01}, \textbf{ALL\_R02}, and \textbf{ALL\_R03}, which establish baseline expectations for anonymization, data sovereignty, and compliance validation. The remaining requirements map to specific use cases as follows: The \textbf{law enforcement} use case maps to \textbf{LAGO\_R01}, \textbf{evidence-based medicine} and \textbf{regional planning} use cases map to \textbf{TEADAL\_R01} and \textbf{TEADAL\_R03}, \textbf{financial governance} and \textbf{smart viticulture} use cases are linked to \textbf{TEADAL\_R02} and \textbf{TEADAL\_R04}, and the \textbf{mobility} use case is associated with \textbf{TEADAL\_R04}.

\end{sloppypar}

\begin{table}[htp]
\centering
\caption{Requirements identified for LAGO and TEADAL projects.} \label{table:requirements-elicited}
\renewcommand{\arraystretch}{1.3} % Adjust row spacing

\fontsize{8pt}{8pt}\selectfont
\begin{tabular}{|c|p{9.2cm}|}
\hline
\textbf{Requirement ID} & \textbf{Requirement Description} \\
\hline
\textbf{ALL\_R01} & Ensure anonymization and secure generation of data to protect sensitive information \\
\hline
\textbf{ALL\_R02} & Allow data owners to retain control over original data to uphold data sovereignty, avoiding off-premises exposure during SDG \\
\hline
\textbf{ALL\_R03} & Implement compliance checks to verify the validity of synthetic data for utility and privacy before sharing it with external entities \\
\hline
\textbf{LAGO\_R01} & Ensure conformance of synthetic data to legal and regulatory frameworks for non-identifiable data, supporting compliance \\
\hline
\textbf{TEADAL\_R01} & Utilize standard data models and pipeline formats for interoperability across multiple organizations and environments \\
\hline
\textbf{TEADAL\_R02} & Provide structured, standard workflows for managing cross-organizational data sharing, particularly in multi-cloud setups \\
\hline
\textbf{TEADAL\_R03} & Ensure confidentiality and control over data prior to sharing, allowing data owners to review and validate SDG outputs \\
\hline
\textbf{TEADAL\_R04} & Design scalable and modular synthetic data workflows that support large-scale data generation and flexible deployment \\
\hline
\end{tabular}%
\end{table}

\begin{figure}[hb] % * makes it span both columns
\centering
\includegraphics[width=0.8\textwidth]{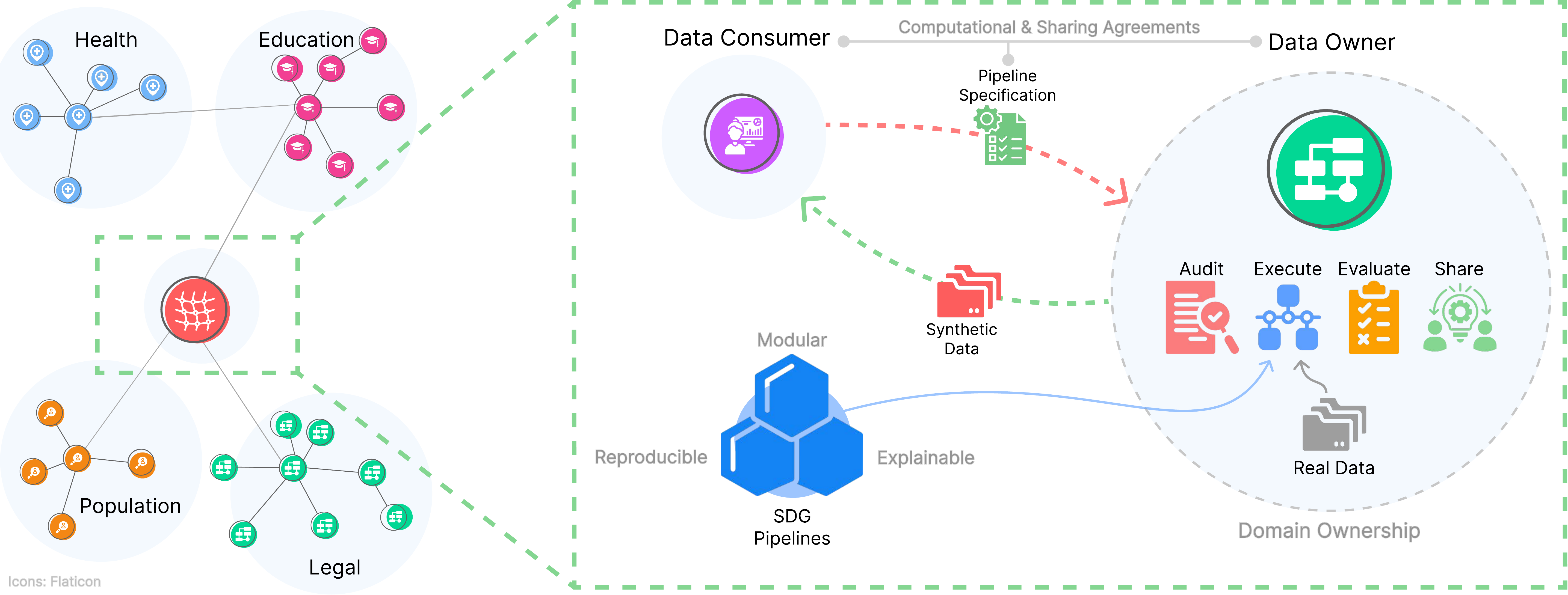} 
\caption{SDG architectural approach, targeting Data Mesh principles of domain  ownership, data sovereignty, and computational governance.}
\label{fig:approach}
\end{figure}

With the requirements defined, the following subsections present a data-sharing architectural approach designed to meet them. Drawing from Data Mesh principles, it emphasizes domain ownership, computational governance, and data sovereignty~\cite{machado2022data,hummel2021data} (see Fig.\ref{fig:approach}). This model shifts SDG toward a more autonomous and programmable framework, aligning with data privacy regulations and the growing need for secure yet scalable synthetic data sharing~\cite{EU2022,EU2023}.

\subsection{Computational Governance for Data Availability}

\begin{sloppypar}

In sensitive domains like healthcare, law enforcement, and finance, computational placement is crucial to ensure data availability without compromising sovereignty~\cite{hummel2021data}. SDG processes must remain localized within clearly defined computational domains to keep data under the jurisdiction of its rightful owner. As shown in Fig.\ref{fig:approach}, real data stays within the data owner's domain, supporting \textbf{ALL\_R01} and \textbf{ALL\_R02}, which call for secure, owner-controlled SDG without off-premises exposure. This localized model also supports \textbf{LAGO\_R01} by promoting compliance and trustworthiness, embedding privacy and transparency mechanisms throughout the SDG pipeline. Computational placement further introduces enhanced privacy guarantees—such as controlled data flows, tailored trust models, and reduced network exposure~\cite{donta2023exploring}, reinforcing \textbf{TEADAL\_R03}, which prioritizes confidentiality and pre-sharing validation.

\end{sloppypar}

Beyond control and privacy, this SDG model promotes a flexible, programmable approach that allows data owners to adapt workflows to evolving needs while retaining oversight~\cite{machado2022data,goedegebuure2024data}. As hybrid and heterogeneous SDG methods grow in demand, especially in AI-driven domains, pipelines must support complex operations, massive datasets, and diverse resources~\cite{bisong2019kubeflow,zhou2020towards,tatineni2021ai}. This requires standardized tools and processes~\cite{zielke2020artificial,hechler2020operationalization}, encompassing both the SDG method stack and the infrastructure stack (pipeline specifications, workflow, and orchestration engines). Such standardization is essential for enabling audits during the development, engineering, and deployment of SDG tools and infrastructure, as well as for meeting industry demands for reliability and trustworthiness~\cite{britotrustops} — core to \textbf{TEADAL\_R02}'s emphasis on structured, cross-organizational sharing, particularly in multi-cloud setups.

Flexible deployment models are also needed to meet sovereignty and compliance goals. As with data mesh principles, deployments should reflect domain ownership and computational governance~\cite{machado2022data}, ideally operating near the original data source under the data owner's oversight~\cite{Majeed23}. This includes all SDG stages, from specification to evaluation. By balancing on-premises and cloud resources, owners can scale workflows while maintaining strict data flow controls, minimizing exposure and maximizing data control and privacy~\cite{beckman2020harnessing,Majeed23,donta2023exploring}. Such setups further strengthen \textbf{ALL\_R01}'s aim to secure and localize data generation.

Reproducibility and explainability are also essential to ensure that SDG processes remain modular, reliable, and auditable across development and deployment. Localized control paired with modular design supports transparent workflows and simplifies audits, helping meet both ethical and legal standards~\cite{hummel2021data,goedegebuure2024data}. This modularity allows individual workflow components to be examined for compliance and performance, important for satisfying \textbf{ALL\_R03}, which requires pre-sharing validation. It also streamlines implementation by enabling separate execution, testing, and validation phases~\cite{dessalk2020scalable}. These steps of auditing, executing, evaluating, and sharing synthetic data, under the data owner domain, are depicted in Fig.~\ref{fig:approach}. 

Furthermore, standardized, reusable SDG components further support cross-domain adoption, reduce redundancy, and simplify integration of new data sources and techniques. This focus on reproducibility and explainability allows pipelines to evolve with changing regulatory and ethical requirements. As privacy demands increase, the ability to validate and explain synthetic data processes becomes critical for maintaining compliance and building trust~\cite{britotrustops}. Together, these practices support the emergence of a sustainable and accountable synthetic data ecosystem that prioritizes sovereignty and responsible sharing~\cite{goedegebuure2024data,hummel2021data}.

\subsection{Programmability and Composability in Scaling SDG Pipelines}

\begin{sloppypar}
As SDG matures, organizing modular mechanisms into structured, programmable settings becomes essential. These should be categorized by function and implemented in cross-platform languages that ensure portability. This enables SDG modules to be delivered declaratively, reliably, and reproducibly, while integrating smoothly across environments, fulfilling \textbf{TEADAL\_R02} and \textbf{TEADAL\_R04} requirements for structured, modular, and scalable workflows.
\end{sloppypar}

Python and R dominate as core tools for implementing SDG libraries~\cite{nagpal2019python,pezoulas2024synthetic}. To support broader integration, individual modules must be transformed into portable components with consistent behaviour across environments. This allows local development and validation, followed by reliable deployment. It also facilitates auditing, as code and execution paths can be reviewed for compliance with security and reliability standards~\cite{shi2021experience}, potentially in automated fashion. Fig.~\ref{fig:approach} illustrates these ideal properties when pipelines are executed in the data owner's domain.

In practice, Docker (for containerization) and Nix (for reproducible configuration) are widely adopted to package and manage components at scale~\cite{dessalk2020scalable,malka2024reproducibility}. These tools offer modularity, isolation, and reproducibility, aligning with scalable workflow practices. Kubernetes adds declarative orchestration, enabling modular, composable services. When paired with workflow engines~\cite{bisong2019kubeflow,dessalk2020scalable}, SDG pipelines are executed as directed acyclic graph (DAG) workflows with clear task dependencies and conditional execution. This structure allows for automated transformation of data, with each step carefully organized and controlled. Again, these choices help target \textbf{TEADAL\_R02} and \textbf{TEADAL\_R04}, in terms of the modularity, interoperability, and scalability of SDG workflows.

Advanced orchestration is also achievable using integrated technologies. Tasks can be labelled and scheduled to run on specific nodes~\cite{oleghe2021container}, or securely offloaded to trusted execution environments (TEEs) in untrusted domains, realizing the concept of privacy preserving data pipelines~\cite{britotrustops}. These TEEs provide secure enclaves for privacy-preserving computation under resource-richer conditions~\cite{KammBBO23}. Through these workflow engines, SDG pipelines can adopt a declarative meta-computation approach that converts pipeline specifications into shareable, auditable artifacts, thus enabling secure, transparent synthetic data sharing across domains.

\subsection{Sharing SDG Pipeline Specifications}

At the core of this synthetic data sharing model is the interaction between data owners (or domain experts, possibly involving data processors) and synthetic data consumers~\cite{eichler2022data,goedegebuure2024data}, now facilitated through a standardized SDG pipeline specification. As shown at the top of Fig.~\ref{fig:approach}, this specification is the output of computational and sharing agreements between both parties. It serves as a declarative, shareable artifact and aligns with \textbf{TEADAL\_R01} by enabling interoperability through a standardized pipeline format. The specification includes a DAG outlining task orchestration, along with resource allocations, execution settings, and code for each SDG stage. Therefore, it becomes the central, controlled artifact around which data-sharing is organized~\cite{eichler2022data,goedegebuure2024data}.

Nonetheless, before implementation, data owners and consumers should clarify data requirements, sensitivity levels, and regulatory constraints~\cite{eichler2022data}. Designated experts can then implement, test, and validate the pipeline to ensure it meets privacy, reliability, and trustworthiness requirements~\cite{rambertimplications}, before handing it off to the data owner. This process allows the owner to review and approve all components before introducing sensitive data. These steps reinforce \textbf{ALL\_R02}, \textbf{ALL\_R03}, \textbf{LAGO\_R01}, and \textbf{TEADAL\_R03}, which emphasize confidentiality, compliance, and control over data prior to sharing.

For data owners, an automated deployment setup enhances sovereignty and simplifies operations. The SDG pipeline is executed within the owner’s jurisdiction to comply with data locality and sovereignty mandates~\cite{hummel2021data,EU2023}. Automating the full process, from specification to result validation, reduces overhead and promotes accessible, auditable SDG practices.

This standardized, portable approach also enables future synthetic data markets, supporting secure cross-sector and cross-organization sharing~\cite{eichler2022data,EU2022,EU2023}. Such a marketplace would allow collaboration while maintaining control over pipeline execution and output validation. As synthetic pipelines become standardized~\cite{zielke2020artificial}, organizations can use these artifacts to build interoperable and secure ecosystems. This aligns with \textbf{TEADAL\_R01} and \textbf{TEADAL\_R02}, which promote structured workflows for trusted data sharing. Ultimately, pipeline specifications can serve as repeatable, trustable artifacts that foster responsible, scalable synthetic data collaboration across domains and borders~\cite{abs-1710-08874,EU2022,EU2023}.

\section{The SynthGuard Framework}\label{sec:synthguard}

\begin{figure*}[htbp] % * makes it span both columns
\centering
\includegraphics[width=0.8\textwidth]{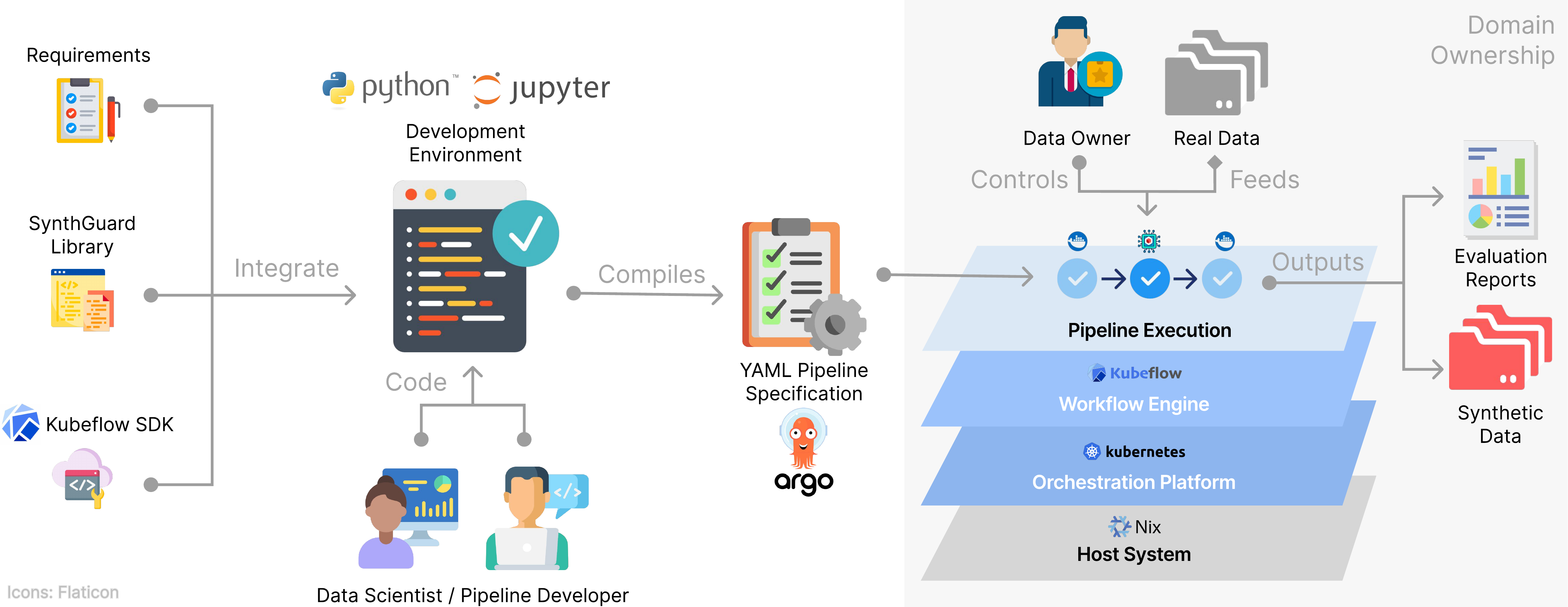} 
\caption{SynthGuard framework overview. The pipeline specification is constructed from modular SDG components (left), compiled and orchestrated through Kubeflow/Kubernetes (centre), and executed under the data owner's control (right). Outputs include synthetic datasets and their corresponding evaluation reports.}
\label{fig:synthguard}
\end{figure*}

Building on the architectural principles discussed, SynthGuard offers a structured SDG workflow that supports data sovereignty, scalability, and privacy. It translates the theoretical foundations into a functional system applicable across diverse use cases. This section details the framework’s core components: actor interactions, requirements elicitation, pipeline specification, deployment models, and evaluation mechanisms. The final subsection presents SynthGuard’s validation and Table~\ref{table:requirements} finally summarizes how its features map to the requirements in Table~\ref{table:requirements-elicited}. Fig.~\ref{fig:synthguard} supports the explanations.

\subsection{Process Overview and Actor Interactions}

SynthGuard structures SDG around a standardized, shareable pipeline specification, keeping computation and orchestration fully under the data owner's control. As shown in the centre of Fig.~\ref{fig:synthguard}, this specification acts as a blueprint containing attributes, generation logic, and evaluation components, guiding collaboration between data owners and consumers.

Data owners, custodians of the original sensitive data, define requirements for synthetic data types, privacy, utility, and deployment models. These inform the construction of an SDG pipeline that aligns with organizational goals and regulatory standards. SynthGuard translates these inputs into a standard specification, jointly agreed upon by both parties and compatible with its workflow engine. This allows the data owner to execute the pipeline independently, ensuring full control over data handling and computation.

Data consumers, such as researchers or analysts, collaborate by specifying desired statistical or causal properties of the synthetic data. SynthGuard supports this through a modular library of SDG components, enabling tailored pipeline assembly. The resulting specification is portable and reproducible, allowing data owners to audit and execute it in a secure environment. Positioned as the central artifact in the data-sharing interaction, the specification governs synthetic data generation while preserving oversight for data owners and meeting consumers’ analytical needs in a transparent, compliant manner.

\subsection{SDG Requirements Elicitation}

Requirements elicitation is a foundational step in SynthGuard, defining parameters to ensure synthetic data aligns with intended use cases, privacy needs, and compliance constraints. It involves specifying the synthetic data type, evaluation preferences, and deployment environment—providing a structured approach for both data owners and consumers. The type of synthetic data is selected based on its intended application, with options including RSD, CSD, ASD, and HSD:

\begin{itemize} 

\item \textbf{RSD:} Realistic synthetic data replicates the statistical structure of real data, preserving distributions, correlations, and other key patterns. It requires access to original datasets.

\item \textbf{CSD:} Causal synthetic data maintains causal relationships and depends on both real data and known causal structures.

\item \textbf{ASD:} Artificial synthetic data is generated from aggregate statistics or predefined rules, without needing raw data. These inputs may also carry privacy constraints.

\item \textbf{HSD:} Hybrid synthetic data combines aspects of RSD, ASD, and/or CSD, offering flexibility to address diverse analytical or policy goals.

\end{itemize}

Evaluation preferences focus on utility and privacy assessments and guide SDG pipeline configuration to ensure outputs meet required standards:

\begin{itemize} 

\item \textbf{Utility evaluation:} Measures how effectively synthetic data substitutes real data in analysis tasks, using univariate, bivariate, or multivariate assessments. This ensures that the synthetic data is useful for intended applications.

\item \textbf{Privacy evaluation:} Evaluates the extent to which re-identification or disclosure risks are mitigated, supporting compliance with privacy regulations.

\end{itemize}

The deployment model defines security levels and resource configurations. Data owners choose between on-premises or cloud-based systems, and container- or VM-based isolation, depending on infrastructure and privacy requirements. On-premises deployment maximizes data control, while cloud environments offer resource scalability. VM-based isolation provides strong separation between tasks, reducing breach risk, while containers offer lightweight, efficient deployment for scalable workloads. SynthGuard’s flexible model—including future support for VM-based TEEs—allows data owners to tailor pipeline isolation and infrastructure, enabling privacy, security, and scalability according to context. We clarify that SynthGuard does not define new SDG models, but rather integrates user-specified methods, such as ML-based CTGAN or rule-based generator, into modular pipelines tailored to the data context.

\subsection{Pipeline Specification, Deployment and Execution}

The first SynthGuard prototype is built on Nix~\cite{nix}, Kubernetes (via Minikube) \cite{kubernetes,minikube}, and Kubeflow Pipelines~\cite{kubeflow-pipelines}, with Argo Workflows as the pipeline specification standard~\cite{argo-workflows}. As shown on the right side of Fig.~\ref{fig:synthguard}, these components form the SynthGuard execution environment, which data owners configure and deploy locally. Nix enables reproducible, cross-platform deployment with isolated dependencies, ensuring a consistent stack across environments.

The pipeline specification begins with the SynthGuard Python library — a set of modular SDG components. Users (data owners, scientists, or developers) compose workflows by selecting and configuring modules tailored to their needs. Using the Kubeflow SDK, these are assembled into executable pipelines, often within a Python Notebook. Once tested, pipelines are exported as Argo Workflows YAML specifications. The left side of Fig.~\ref{fig:synthguard} depicts these concepts. This specification serves as a portable artifact which data owners can audit and review before running it against sensitive data.

Data owners then deploy these specifications within their SynthGuard stack. Kubeflow orchestrates execution, monitoring, and adjustment of components, while Kubernetes handles resource allocation and scaling, supporting both single- and multi-node clusters as needed. After execution, synthetic data and its corresponding privacy and utility reports—depicted on the right of Fig.~\ref{fig:synthguard}—are produced directly by the pipeline. These reports help verify compliance with privacy standards and assess data utility. If validated, the synthetic dataset can be safely shared with data consumers, completing the SDG process under controlled conditions.

\subsection{Privacy and Utility Evaluation Mechanisms}

SynthGuard integrates evaluation mechanisms directly into SDG pipelines to balance diagnostic accuracy, utility, and privacy. Diagnostic evaluation includes essential checks to ensure synthetic data meets quality standards in format, validity, and structure~\cite{PatkiWV16}. For example, validity checks confirm continuous values fall within the observed real-data range, and categorical columns retain valid entries. Structural checks ensure consistency in column names and data types.

Utility evaluation measures how well synthetic data replicates the statistical properties of the original. \textit{Univariate} comparisons assess marginal distributions, while \textit{bivariate} evaluations capture correlations between columns. At the population level, propensity score–based metrics provide deeper insight~\cite{PatkiWV16,dankar2021fake}. Key metrics include observed propensity score mean-squared error (observed pMSE), which measures how closely synthetic data conforms to the propensity scores of real data; null-standardized pMSE (standard pMSE), which compares the observed pMSE to a distribution derived from a null model; observed-null pMSE ratio, which indicates how closely the observed pMSE matches the null model; and Kolmogorov-Smirnov distance (SPECKS), which quantifies the difference between the cumulative distribution functions of the propensity scores of real and synthetic data. 

Privacy evaluation is essential in sensitive or regulated environments. Metrics assess risks of synthetic data revealing sensitive information~\cite{PatkiWV16,el2020practical}. CategoricalCAP estimates risks from inference attacks using public and synthetic data. NewRowSynthesis flags exact or near matches between real and synthetic rows. Inference attack scores measure the likelihood of attribute inference via statistical or ML models. Target correct attribution probability (TCAP) evaluates the probability of matching sensitive values based on key attributes. Additional metrics like minimum nearest-neighbour and sample overlap scores assess privacy risk based on proximity between real and synthetic data\footnote{CategoricalCAP and NewRowSynthesis are from SDV: \url{https://docs.sdv.dev/sdv}; other metrics are from SynthGauge: \url{https://datasciencecampus.github.io/synthgauge/index.html}.}.

\section{Validation}\label{sec:validation}

To validate SynthGuard, we followed an iterative process combining expert feedback, multi-environment deployments, and SDG quality and privacy assessment. Co-developed with domain experts from both projects, SynthGuard was refined through ongoing feedback from workshops and collaborative sessions, aligning with legal, technical, and operational requirements (Table~\ref{table:requirements-elicited}). It was tested in three deployment settings: a \textbf{local} environment for testing and debugging, an \textbf{on-premises} setup to verify data sovereignty compliance, and a \textbf{cloud-based} configuration to assess scalability and distributed workflow integration.

Validation consisted of implementing SDG pipelines within the framework, executing them with varied inputs (e.g., real data, schemas, distributions) across environments, and verifying compliance through the generation of utility and privacy evaluation reports. Following the workflows described in Section~\ref{sec:approach}, we confirmed SynthGuard's modularity, flexibility, and reproducibility, and validated pipeline execution using its modular libraries and orchestration tools.

For both projects, we generated diverse datasets (RSD, ASD, HSD) using ML-based methods like CTGAN~\cite{PatkiWV16}, depending on data availability. For ASD, rule-based synthesis was used due to the lack of real data. Privacy and utility evaluations were integrated into pipeline execution, comparing distributions and structures to ensure representativeness and minimize disclosure risk. For ASD, direct evaluations were not conducted due to the absence of real data, though rule-based leakage checks may be explored in future work to assess privacy risks.

\begin{figure}[htbp]
    \centering
    \includegraphics[width=0.7\linewidth]{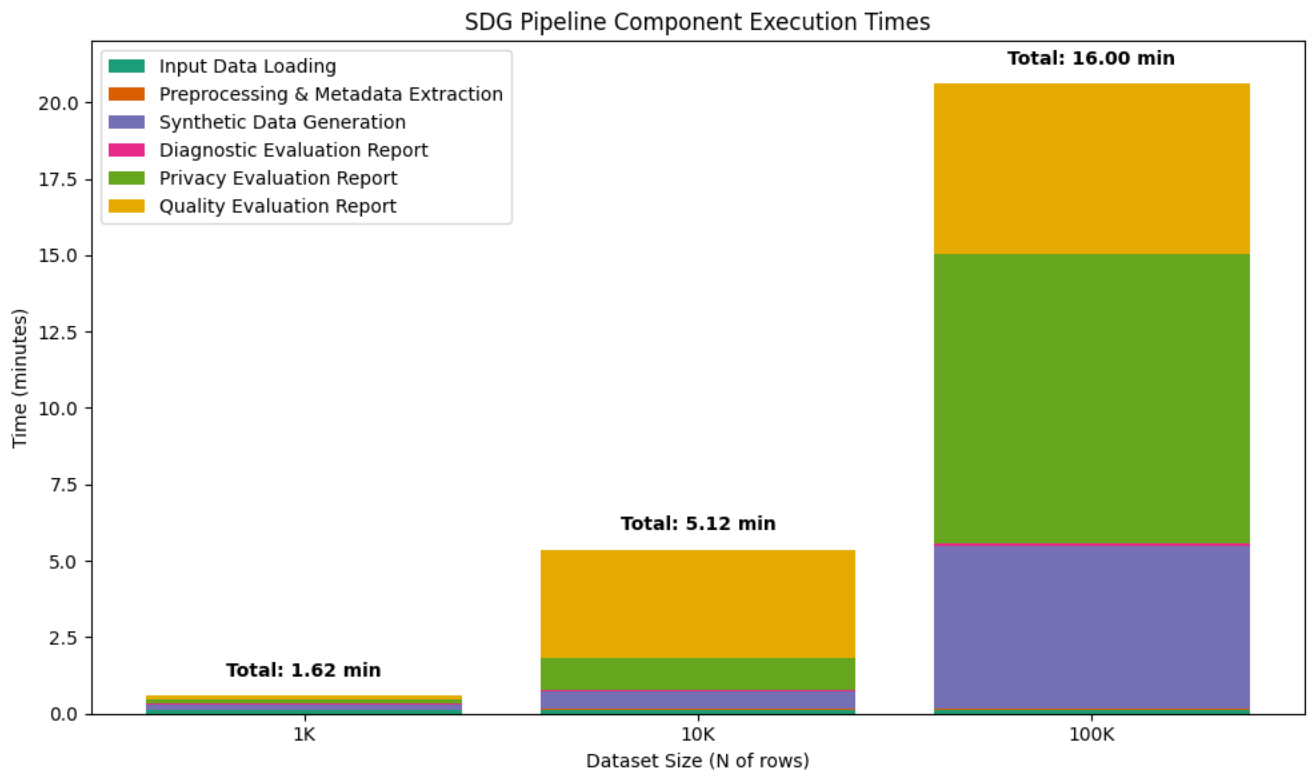}
    \caption{Execution time per SDG pipeline component across dataset sizes.}
    \label{fig:sdg_runtime}
\end{figure}

To assess scalability, we benchmarked SynthGuard on \textbf{law enforcement} use case datasets of 1K, 10K, and 100K rows using a cloud VM (8 vCPUs, 12GB RAM). While the other use cases informed the architectural requirements, the validation scope focused on pipeline execution and scalability, which were sufficiently demonstrated using one use case. The pipeline included data loading, preprocessing, generation, and parallel evaluation (quality, diagnostic, privacy). As shown in Figure~\ref{fig:sdg_runtime}, loading and preprocessing remained constant ($\sim$0.1 min), while generation rose from 0.14 to 5.35 min, privacy from 0.14 to 9.46 min, and quality from 0.10 to 5.59 min. Total runtime scaled sublinearly: 1.6 min (1K), 5.1 min (10K), and 16 min (100K). At 100K rows, privacy and quality evaluations accounted for over 90\% of runtime, but with parallel execution via Kubeflow, their durations didn’t accumulate. These results confirm SynthGuard’s scalability and show how modular, concurrent design mitigates bottlenecks at scale.

\begin{figure*}[htbp] 
\centering 
\includegraphics[width=0.9\textwidth]{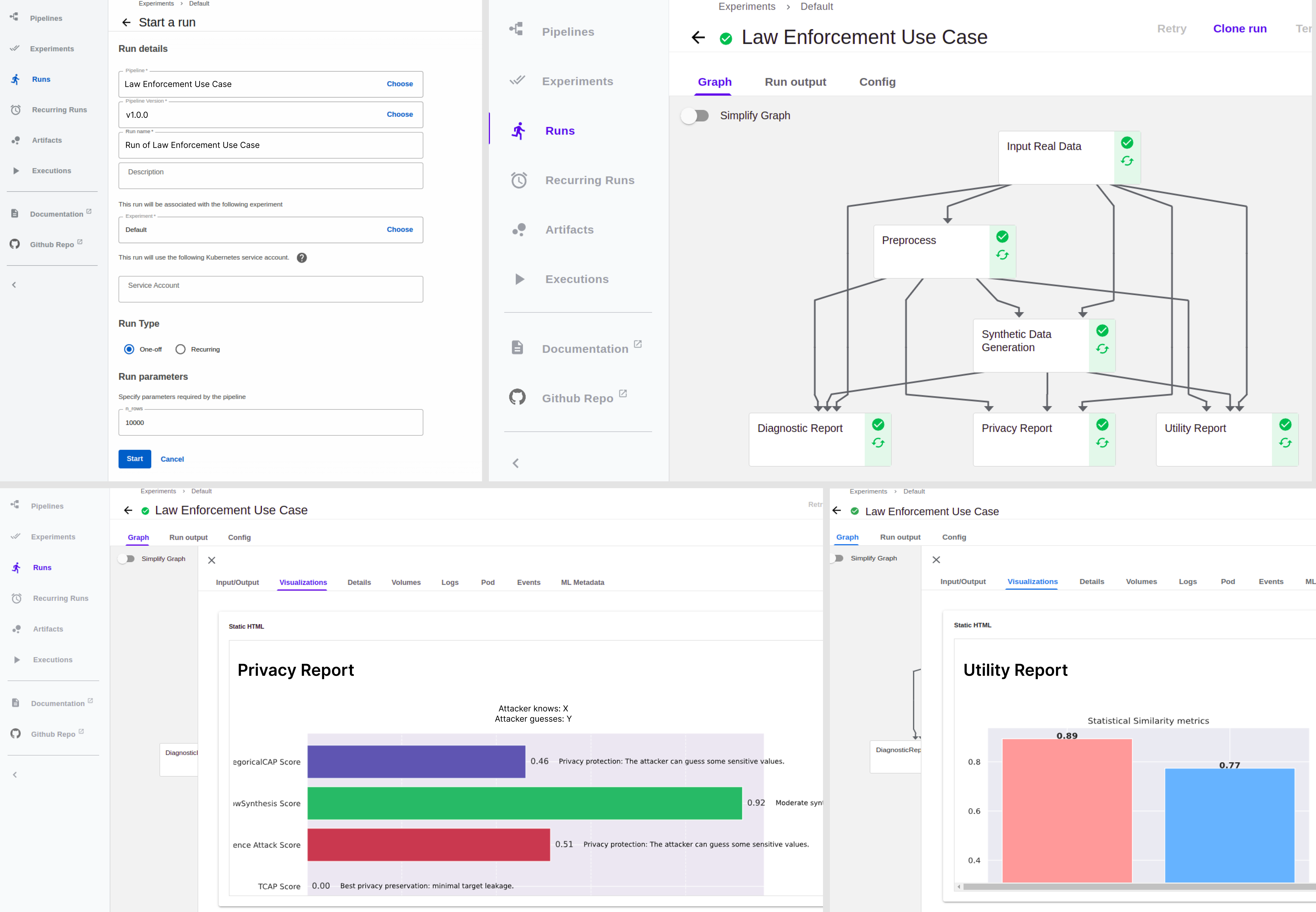} 
\caption{Top: Example of a law enforcement SDG pipeline visualized via Kubeflow. Bottom: Privacy and utility reports accessible per pipeline component.} 
\label{fig:example-law-enforcement-use-case} 
\end{figure*}

Figure~\ref{fig:example-law-enforcement-use-case} illustrates a pipeline execution in the \textbf{law enforcement} use case. The pipeline specification is first loaded into the system, and the data owner provides input parameters, such as the target dataset size. The pipeline then loads the real data, generates the synthetic data, and compiles reports based on the chosen SDG method. Arrows represent the sequential flow and output transfers between components. Users can view execution code, logs, and download datasets and reports through the same interface.

Table~\ref{table:requirements} summarizes how SynthGuard meets project-specific requirements. The validation process confirmed SynthGuard's feasibility, adaptability, and robustness across varied environments. Future validation work includes additional performance benchmarking, large-scale deployment validation, and further automation of compliance checks.

{\small
\begin{table}[htp]
\centering
\caption{LAGO and TEADAL requirements fulfilled by SynthGuard.} \label{table:requirements}

\renewcommand{\arraystretch}{1.3} % Adjust row spacing

\fontsize{8pt}{8pt}\selectfont
\begin{tabular}{|c|p{9cm}|}
\hline
\textbf{Requirement ID} & \textbf{SynthGuard Features} \\ \hline

\textbf{ALL\_R01} & Provides on-site SDG with adaptable privacy-preserving mechanisms \\ \hline

\textbf{ALL\_R02} & Enables local SDG without external transfer \\ \hline

\textbf{ALL\_R03} & Includes utility and privacy validation in workflows \\ \hline

\textbf{LAGO\_R01} & Enables the programmability of legally compliant SDG mechanisms \\ \hline

\textbf{TEADAL\_R01} & Supports standard pipeline specifications \\ \hline

\textbf{TEADAL\_R02} & Enables modular, reproducible pipelines, with flexible deployment models \\ \hline

\textbf{TEADAL\_R03} & Enables local and secure output validation by data owners before sharing \\ \hline

\textbf{TEADAL\_R04} & Provides modular components for SDG and flexible scaling \\ \hline

\end{tabular}%
\end{table}
}

\section{Conclusion and Future Work} \label{sec:conclusion}

Privacy-enhancing technologies (PETs) are increasingly adopted to balance data utility and privacy in data-driven applications. Among them, synthetic data generation (SDG) enables representative datasets without compromising individual privacy. However, the need for secure, privacy-preserving, and scalable SDG frameworks that uphold data sovereignty and regulatory compliance, without relying on centralized processing, remains critical.

To address this, we propose SynthGuard, a modular, computational governance–oriented framework for privacy-preserving and scalable SDG. Developed iteratively with domain expert input and validated in real-world scenarios, we demonstrate through SynthGuard a balanced approach to privacy, security, and compliance, with support for SDG pipelines and utility and privacy assessments in diverse environments.

While effective in orchestrating modular pipelines, SynthGuard currently targets single-table scenarios and selected use cases. Future work includes support for multi-table relational data, integration of compliance validation, and deployment with secure multi-party computation and trusted execution environments~\cite{KammBBO23}. We also plan to adopt dataspace protocols~\cite{atzori2024dataspaces,alsamhi2024empowering}, expand compatibility with mainstream infrastructure, and explore automated workflow generation and performance optimization via GPU/TPU acceleration. As data-sharing ecosystems grow in complexity, SynthGuard will evolve with reusable pipelines, real-time generation, and regulatory alignment.

\bibliographystyle{splncs04}
\bibliography{mybibliography}

\end{document}